\documentstyle[11pt,twoside,jltp,psfig]{article}
\newcommand{\beq}{\begin{equation}}
\newcommand{\eeq}{\end{equation}}
\newcommand{\beqa}{\begin{eqnarray}}
\newcommand{\eeqa}{\end{eqnarray}}
\newcommand{\ba}{\begin{array}}
\newcommand{\ea}{\end{array}}

\title{Bosonic clouds with attractive interaction 
beyond the local interaction approximation}

\author{L. Reatto$^{1}$, A. Parola$^{1,2}$ and 
L. Salasnich$^{1}$
\address{$^{1}$INFM and Dipartimento di Fisica, Universit\`a di Milano, \\
Via Celoria 16, 20133 Milano, Italy \\
$^{2}$Istituto di Scienze Fisiche, Universit\`a di Milano, \\
Via Lucini 3, Como, Italy}}

\runninghead{L. Reatto, A. Parola and L. Salasnich}
{Bosonic clouds beyond the local interaction approximation}
       
\begin{document}

\begin{abstract}
We study the properties of a Bose--Einstein condensed cloud of atoms 
with negative scattering length confined in a harmonic trap. 
When a realistic non local (finite range) effective interaction is taken into
account, we find that, besides the known low density metastable solution,
a new branch of Bose condensate appears at higher density.
This state is self--bound but its density can be quite low if 
the number $N$ of atoms is not too big. 
The transition between the two classes of solutions as a function
of $N$ can be either sharp or smooth according to 
the ratio between the range of the attractive interaction 
and the length of the trap. A tight trap leads to a smooth transition. 
In addition to the energy and the shape of the cloud we study 
also the dynamics of the system. 
In particular, we study the frequencies of collective 
oscillation of the Bose condensate 
as a function of the number of atoms both in the local and in the 
non local case. Moreover, we consider the dynamics of the cloud 
when the external trap is switched off. 


\end{abstract}

\maketitle

\section{INTRODUCTION}

In the standard treatment of Bosonic alkali atoms in a trap, 
a local form (i.e. momentum independent) is assumed as effective 
interatomic interaction$^{1}$. This can not be completely correct when
the scattering cross section has a significant momentum dependence
already at very low momenta$^{2}$. 
This is the case of $^{7}Li$, a particularly
interesting case due to its negative scattering length. This
momentum dependence implies$^{3}$ that the effective interaction is non local
changing from attractive to repulsive at a characteristic range $r_e$. 
\par
Recently we have studied$^{4}$ the ground state of $^7Li$ atoms with a 
non local interaction in a harmonic trap $U_{\rm ext}({\bf r})=
{1\over 2}m\omega_0^2 r^2$ and we have shown the existence of a new 
branch of states intermediate in density between the known very dilute 
state and the collapsed high density state. Here we study how the 
non locality affects the dynamics of the system. 
We assume that the attractive potential has a finite range $r_e$ and
in addition we allow for the presence of a repulsive contribution
which is modeled as a {\it local} positive term defined by a ``high energy"
scattering length $a_R > 0$. The effective interaction is then written 
in the following form$^{4}$: 
\begin{equation}
v_{\rm eff}(k)={4\pi\hbar^2\over m}
\left [a_R + (a_T-a_R)\,f(kr_e) \right ] \; ,
\end{equation}
where $f(x)=(1+x^2)^{-1}$. We use interaction parameters 
appropriate for $^7Li$: $a_T=-27\,a_B$, $r_e=10^3\,a_B$ and $a_R=6.6\,a_B$ 
(where $a_B$ is the Bohr radius). 

\section{CONDENSATE GROUND STATE}

The ground state wavefunction of a cloud of $N$ atoms 
is determined  by minimizing the Gross--Pitaevskii (GP) functional 
${\cal E}[\Psi]$, where $\Psi({\bf r})$ is the wavefunction 
of the condensate$^{5}$. 
In the ground state $\Psi({\bf r})$ is positive definite 
and spherically symmetric. 
\par
As a first step, we discuss an approximate variational approach to
this problem which already shows the main features of the exact solution.
As a trial wavefunction we choose a Gaussian 
with a single variational parameter $\sigma$ (standard deviation)  
which defines the size of the cloud in units of the harmonic 
oscillator length $a_H=(\hbar/(m\omega_0))^{1/2}$. With this choice, 
the energy ${\cal E}(\sigma )$ can be analytically expressed in 
terms of elementary functions. The extrema of ${\cal E}(\sigma )$ 
are obtained as solutions of an algebraic equation, which 
gives the number of bosons as a function of the size 
$\sigma_0$ of the cloud: 
\beq 
N =(1-\sigma_0^4)\Big[ -\gamma_R \sigma_0^{-1} -{1\over 3}\tau_1 \sigma_0 
+{2\over 3 \sqrt{\pi}} \chi \tau_2 \sigma_0^3 -{2\over 3} \chi^2 \tau_2 
\sigma_0^4 g(\chi \sigma_0 ) \Big]^{-1} \; , 
\eeq
where $\gamma_R=(2/\pi)^{1/2}a_R/a_H$, 
$\tau_1=(2/\pi)^{1/2}a_H(a_T-a_R)/r_e^2$, 
$\chi=2^{-1/2}a_H/r_e$, $\tau_2=a_H^2(a_T-a_R)/r_e^3$ 
and $g(x) = {\rm erfc}(x) \exp{(x^2)}$ with 
${\rm erfc}(x)=1-{\rm erf}(x)$ the complementary
error function. 
This equation has either one or three positive roots 
depending on the parameters and on number $N$ of atoms in the cloud. 
When three solutions are present, the intermediate one represents 
an unstable state (i.e. a local maximum of the 
energy) while the other two respectively describe a low density metastable 
solution and a minimum which represents the stable solution within 
GP approximation. 
The variational results for three typical trap
sizes are shown in Fig. 1, where the mean radius is plotted 
as a function of $N$. 
For comparison, we also show the radius of the cloud when a local 
interaction is assumed ($r_e=0$ in Eq. (1)). In this case there 
is a critical number $N_c\simeq 0.67 a_H/a_T$ of bosons beyond 
which the cloud collapses$^{1}$. 

\begin{figure}
\centerline{\psfig{file=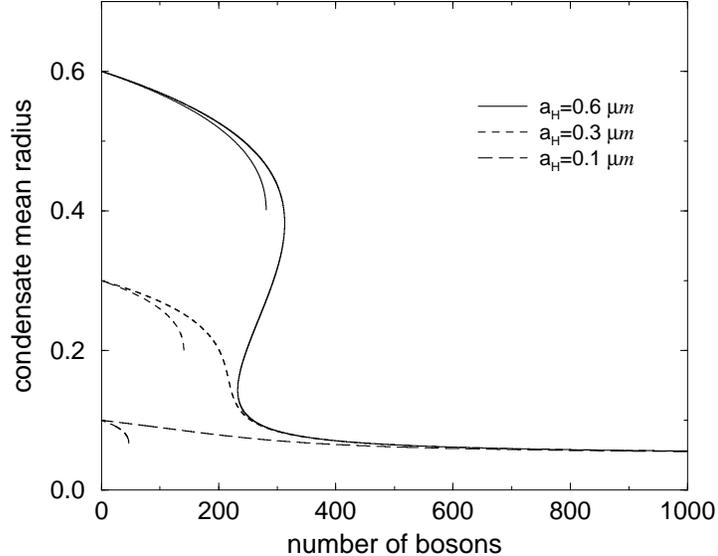,height=3.0in}}
\caption{Mean radius of the condensate, in micron, 
as a function of the number of bosons for 3 different traps. 
The lines with a end point represent the results 
with local interaction.}  
\end{figure}

\par 
We have also computed the exact 
solution of the GP equation, obtained by numerical 
integration of the corresponding self-consistent Schr\"odinger equation. 
The variational approach is always very close to 
the exact solution$^{4}$. 
The effects of non-locality are always important for very tight traps 
while for larger traps non-locality is important just when the radius of 
the cloud rapidly drops for increasing $N$. 
This ``transition" is discontinuous for 
large traps, where the reentrant behavior of the curve shows the
presence of an unstable branch. By reducing the trap size, however, this
discontinuity is strongly reduced and, below about $a_H=0.3\,\mu$m, 
the unstable branch disappears and there is a smooth evolution from a very 
dilute cloud to a less dilute state with an increasing density as $N$ grows. 
\par 
For large $N$ the size of the cloud is remarkably 
independent of the trap size suggesting that the atoms 
are in a self-bound configuration. We have verified this effect 
by integrating numerically the time dependent 
GP equation and by studying the dynamics of the cloud when 
the external trap is switched off. The condensate expands 
when $N<N_{cl}\simeq 234$. For larger $N$ the condensate oscillates 
around the minimum of the energy ${\cal E}(\sigma )$ 
in absence of external trap (see also next section). 
Thus, within our representation of the effective interatomic 
interaction, $234$ is the minimum number needed to get a self-bound 
low density cloud of $^7Li$ atoms. Its average density is about 
$10^{16}$ at/cm$^{3}$. This self-bound state might have a rather 
short life time due to recombination precesses. 

\section{CONDENSATE COLLECTIVE OSCILLATIONS}

The condensate undergoes density oscillations around the 
minimum $\sigma_0$ of the energy ${\cal E}(\sigma )$. 
In the local case, it has been shown that the monopole collective 
oscillation $\omega \to 0$ as $N\to N_c$ with a $1/4$ power law$^{6}$. 
It is interesting to analyze what happens by including non locality. 
We use two schemes: the numerical integration of the time--dependent 
GP equation and an approximate analytical study of the small 
oscillations around the minimum of the energy function of the condensate. 
\par
Following Ref. 6, we associate with the collective motion a kinetic energy 
of the form 
\beq
{\bar T}={1\over 2} m N {\dot r}^2 = 
{3\over 4} {N \hbar \over \omega_0} 
{\dot \sigma}^2 \; , 
\eeq
where $\sigma$ is again the standard deviation of the Gaussian trial 
wavefunction in units of the harmonic oscillator length $a_H$. 
The dynamics of the collective excitations is determined by 
$\bar{T}$ and by the quadratic part of the energy ${\cal E}$ 
expanded in powers of $(\sigma -\sigma_0)$. Some elementary steps 
lead to a remarkably simple expression for the monopole frequency. 
In the local case we find 
\beq
\omega = \omega_0 \Big[5-\sigma_0^{-4}\Big]^{1/2} \; ,
\eeq 
where $\sigma_0$ is related to $N$ by $N=(\sigma_0^5-\sigma_0)/\gamma$ 
with $\gamma=(2/\pi)^{1/2}a_T/a_H$. From Eq. (4) we verify that 
$\omega = 2 \omega_0$ for $\sigma_0=1$, 
and $\omega \to 0$ for $\sigma_0\to 5^{-1/4}$ (i.e. for $N\to N_c$). 
Instead, in the non local case, the frequency reads 
$$ 
\omega  = \omega_0 \Big[ 3\sigma_0^{-4}+1+N 
\Big( 4 \gamma_R \sigma_0^{-5} + 
{2\over 3}\tau_1 \sigma_0^{-3} 
$$
\beq
-{2\over 3}\chi^2\tau_2 g(\chi \sigma_0)(1+2\chi^2\sigma_0^2)
+{4\over 3\sqrt{\pi}}\chi^3\tau_2 \sigma_0 \Big) \Big]^{1/2} \; ,
\eeq 
where $\sigma_0$ is related to $N$ and the interaction parameters 
by Eq. (2). 

\begin{figure}
\centerline{\psfig{file=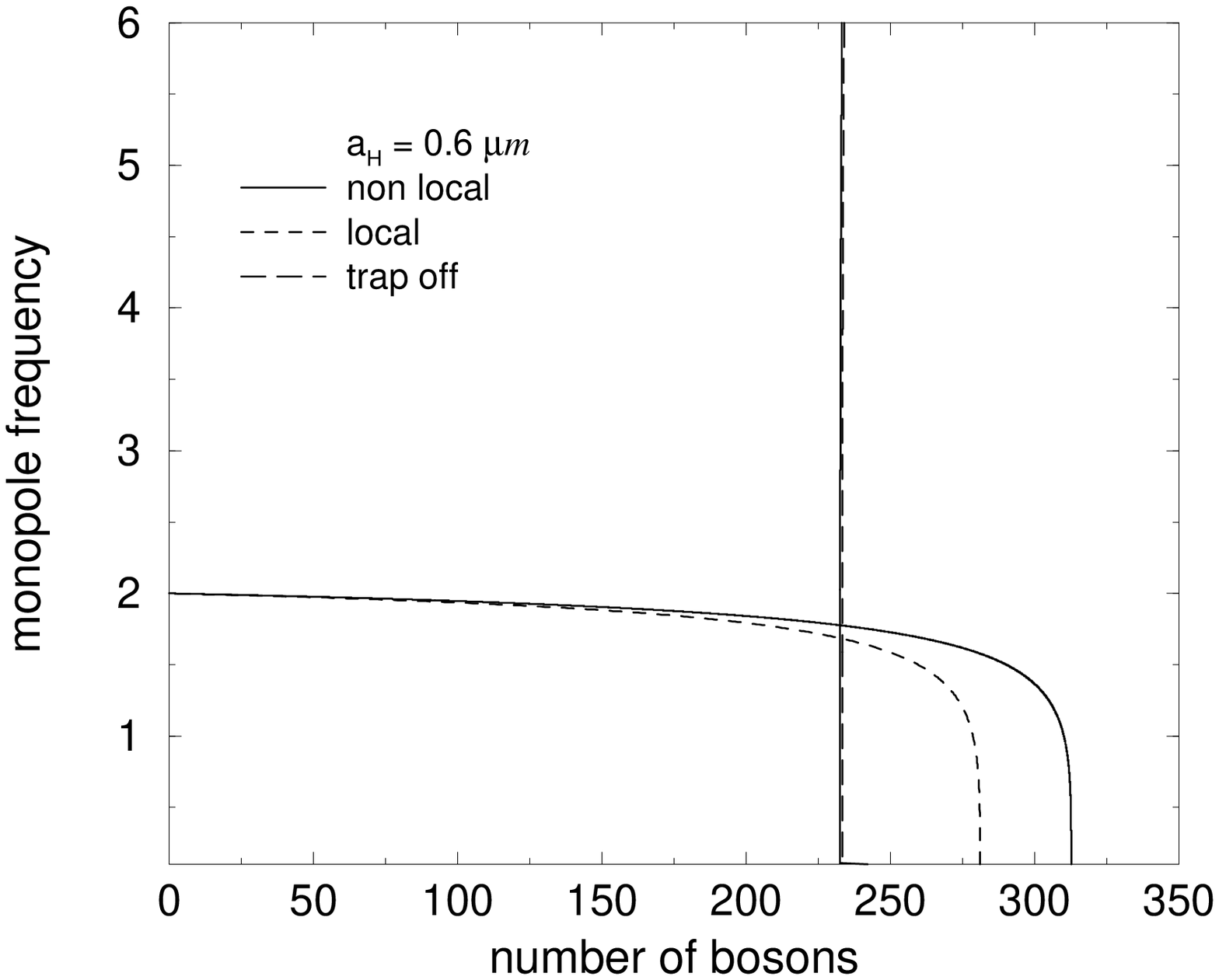,height=2.4in}}
\centerline{\psfig{file=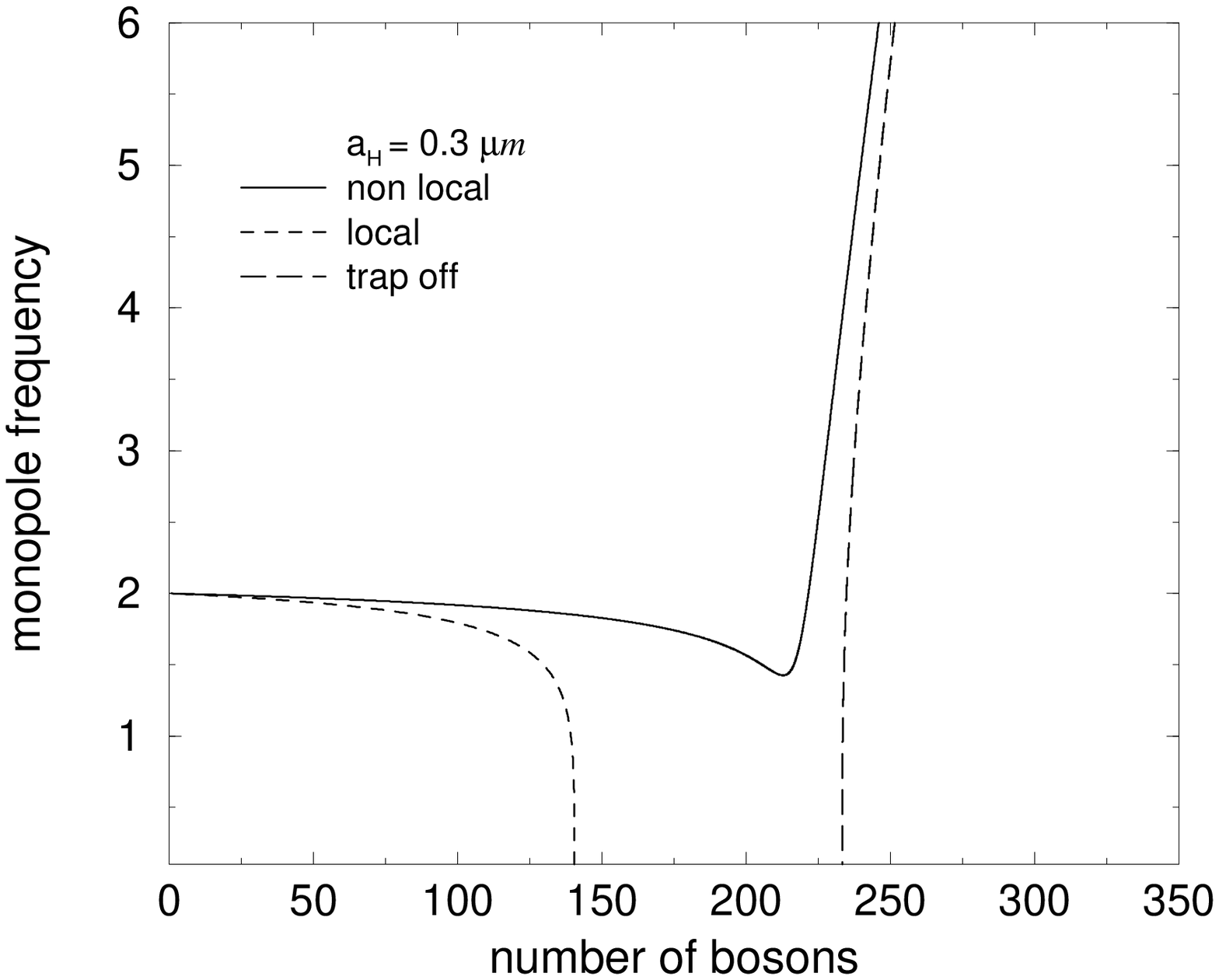,height=2.4in}}
\centerline{\psfig{file=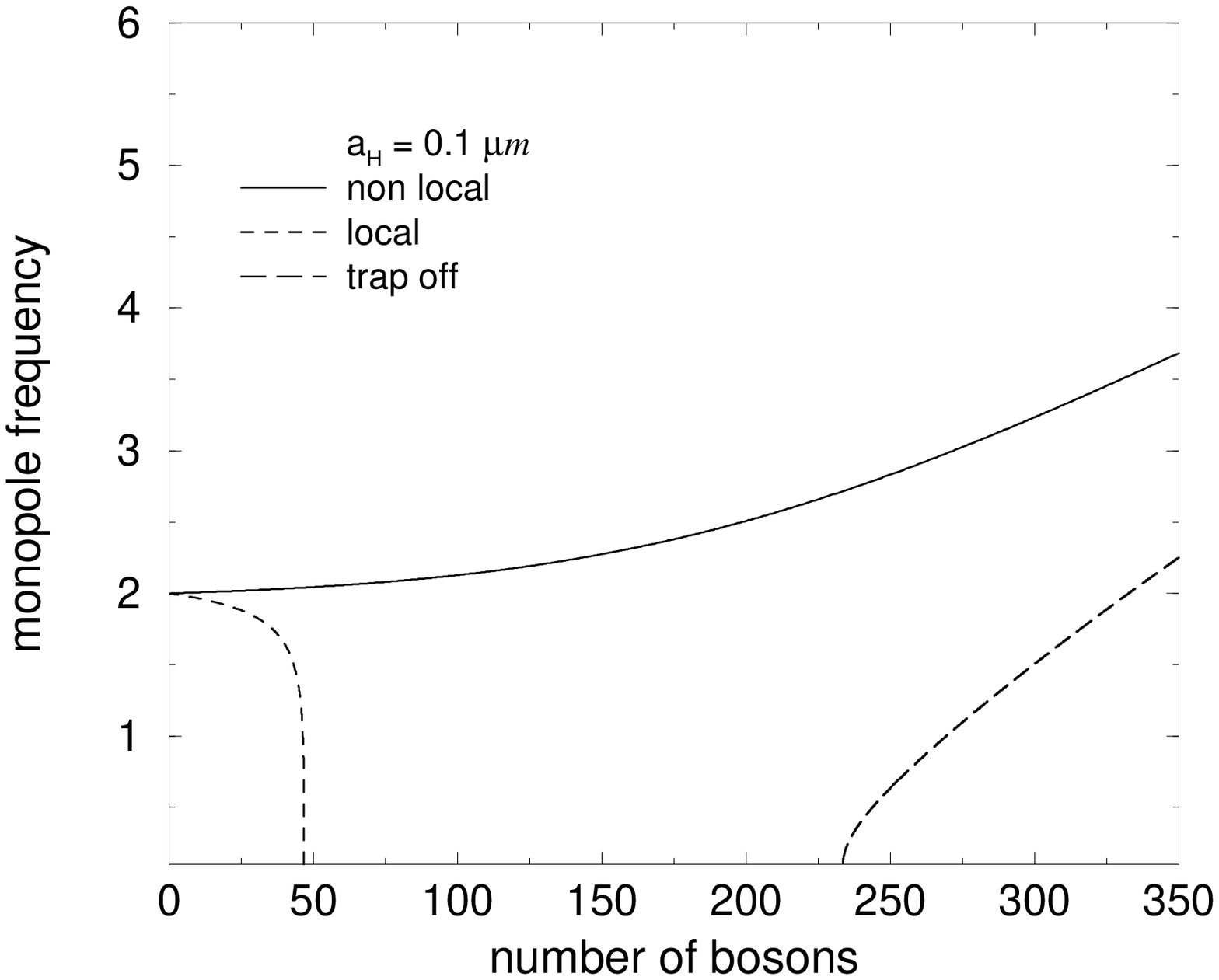,height=2.4in}}
\caption{Monopole frequency of the condensate, in units of trap frequency, 
as a function of the number of bosons for 3 different traps. 
From top to bottom: $a_H=0.6\;\mu$m, $a_H=0.3\;\mu$m, $a_H=0.1\;\mu$m 
(respectively, $\omega_0=25.10$ kHz, $\omega_0=100.41$ kHz, 
$\omega_0=903.67$ kHz). Note that in the trap off case 
there is a well defined frequency only for $N>234$.}  
\end{figure}

In Fig. 2 we show the monopole collective frequency 
of the condensate as a function of $N$ for 3 traps for 
non local, local and trap--off cases. This figure can be easily 
obtained by using Eq. (2), Eq. (4) and Eq. (5). In the non local case, 
for the larger trap ($a_H=0.6\;\mu$m), where there is a 
reentrant behavior,  we see two branches: 
One branch starts from small $N$ and corresponds to the larger cloud. 
The frequency of this branch is very close to the result given by 
the local approximation. In the second branch $\omega$ starts from zero 
at the lowest limit of the reentrant behavior in Fig. 1 and it raises 
rapidly as $N^{1/2}$. For traps of intermediate size ($a_H=0.3\;\mu$m 
in Fig. 2), $\omega$ has a dip in the transition region between 
the low density state and the self bound state. For very small traps, 
there is only one branch and the frequency increases smoothly with the number 
of bosons. As discussed in the previous section, when the external 
trap is switched off the condensate oscillates if $N>N_{cl}\simeq 234$. 
This frequency starts from $0$ at $N_{cl}$ and it 
approaches the trap--on (non local) frequency by increasing $N$. 
As shown in Table 1, there is good agreement 
between the variational monopole frequency of Eq. (5) 
and the numerical one (non local case), 
obtained by solving the time--dependent GP equation. 

\begin{table}[t]
\begin{center}
\footnotesize
\begin{tabular}{|ccc|} \hline\hline 
$N$ & $\omega$ (numerical) & $\omega$ (analytical) \\ 
\hline
100 & 1.91 & 1.92      \\ 
200 & 1.64 & 1.57      \\ 
250 & 6.74 & 6.62      \\
300 & 12.69 & 13.91    \\
500 & 33.96 & 37.25    \\
1000 & 75.70 & 77.85   \\
\hline\hline
\end{tabular}
\end{center} 
\caption{Numerical and analytical monopole frequency, 
in units of the trap frequency, for different values 
of the number $N$ of bosons. $a_H=0.3\;\mu$m.}
\vspace{0.2cm}
\end{table}
 
To conclude, we observe that it should be kept in mind that in the large $N$ 
limit our results are only qualitative because the GP equation 
itself breaks down and interaction effects are expected to produce 
a depletion of the condensate when $\rho |a_T|^3$ is not very small. 

%

\end{document}